\documentclass[aps,prd,twocolumn,showpacs,superscriptaddress,nofootinbib,preprintnumbers]{revtex4-2}

\usepackage[pdftex,colorlinks]{hyperref}
\usepackage[dvipsnames]{xcolor}
\definecolor{phthaloblue}{rgb}{0.0, 0.06, 0.54}

\hypersetup{
    colorlinks=true,
    linkcolor=MidnightBlue,
    citecolor=MidnightBlue,
    filecolor=MidnightBlue,
    urlcolor=MidnightBlue,
    }
\usepackage[pdftex]{graphicx}
\usepackage{bm}
\usepackage{eucal}  
\usepackage{mathrsfs}
\usepackage{cancel}
\usepackage{here}
\usepackage{comment,braket}
\usepackage{epsf}
\usepackage{amsmath}
\usepackage{graphics}
\usepackage{amsfonts}
\usepackage{amssymb}
\usepackage{latexsym}
\usepackage{color}
\usepackage{natbib}
\usepackage[normalem]{ulem}
\input{colordvi.tex}
\usepackage{feynmp}
\usepackage[utf8]{inputenc}
\DeclareUnicodeCharacter{2212}{-}
\usepackage{orcidlink}
\usepackage{multirow}

 \newcommand{\brac}[2]{ \left( \frac{#1}{#2} \right) } 

\newcommand{\be}{\begin{eqnarray}}
\newcommand{\ee}{\end{eqnarray}}

\newcommand{\beq}{\begin{equation}}
\newcommand{\eeq}{\end{equation}}

\newcommand{\Eq}[1]{Eq.~(\ref{#1})}

\begin{document}

\preprint{\tt  FERMILAB-PUB-25-0034-T}

\title{Observing Dark Matter Decays to Gravitons via Graviton-Photon Conversion}

\author{David I. Dunsky}
\email{ddunsky@nyu.edu}
\affiliation{Center for Cosmology and Particle Physics, Department of Physics, New York University, New York, NY, USA}

\author{Gordan Krnjaic\,\orcidlink{0000-0001-7420-9577}}
\email{krnjaicg@fnal.gov}
\affiliation{Theoretical Physics Division, Fermi National Accelerator Laboratory, Batavia, IL, USA}
\affiliation{Department of Astronomy \& Astrophysics, University of Chicago, Chicago, IL USA}
\affiliation{Kavli Institute for Cosmological Physics, University of Chicago, Chicago, IL USA}

\author{Elena Pinetti}
\email{epinetti@flatironinstitute.org}
\affiliation{Center for Computational Astrophysics, Flatiron Institute, New York, NY 10010, USA}
\affiliation{Theoretical Physics Division, Fermi National Accelerator Laboratory, Batavia, IL, USA}
\affiliation{Kavli Institute for Cosmological Physics, University of Chicago, Chicago, IL USA}
\affiliation{Emmy Noether Fellow, Perimeter Institute for Theoretical Physics, 31 Caroline Street N., Waterloo, Ontario, N2L 2Y5, Canada}

\date{\today}

\begin{abstract}
Since dark matter is only known to have gravitational interactions, it may plausibly decay to gravitons on cosmological timescales. Although such a scenario can be easily realized, there are currently no known limits on this possibility based on indirect detection searches. We find that the gravitons produced in dark matter decays can convert to photons in large-scale magnetic fields along the line of sight to an observer. These conversions primarily occur within cosmological filaments which  occupy a large (order unity) volume fraction and
contain $\sim 10-100$ nG fields with $\sim$ Mpc coherence lengths. Thus, dark matter decays to gravitons predict an irreducible population of extragalactic {\it photons}, which we constrain using the extragalactic gamma-ray background measured by the {\it Fermi}-LAT telescope. Using this conservative method, we place the first limits on the dark matter lifetime in the $0.1 {\rm GeV} - 10^8$ GeV mass range, assuming only decays to gravitons. 
We also make  projections for the Advanced Particle-astrophysics Telescope, which can improve sensitivity to this DM decay channel by an order of magnitude beyond those we set using {\it Fermi}-LAT data. 
\end{abstract}

\bigskip
\maketitle

\medskip
\noindent
{\bf Introduction.} 
The evidence for dark matter (DM) is overwhelming, but its only known interaction with the Standard Model (SM) is through gravity. If there is no symmetry to stabilize the DM, a generic possibility is that 
it can decay to {\it gravitons}. Since graviton couplings to the SM are suppressed by powers of the Planck mass, such a scenario is exceedingly difficult to probe using established techniques. Although there are indirect detection limits on DM decays to all other SM particles (including neutrinos), to date, there are no observational limits on the graviton channel.\footnote{However, there are generic bounds on DM decays to invisible radiation from probes of large scale structure, which require that the DM comoving mass density be constant  to within $\sim 4$ \% over the age of universe \cite{Poulin:2016nat}.
 }

\begin{figure}[t]
\includegraphics[width =\linewidth]{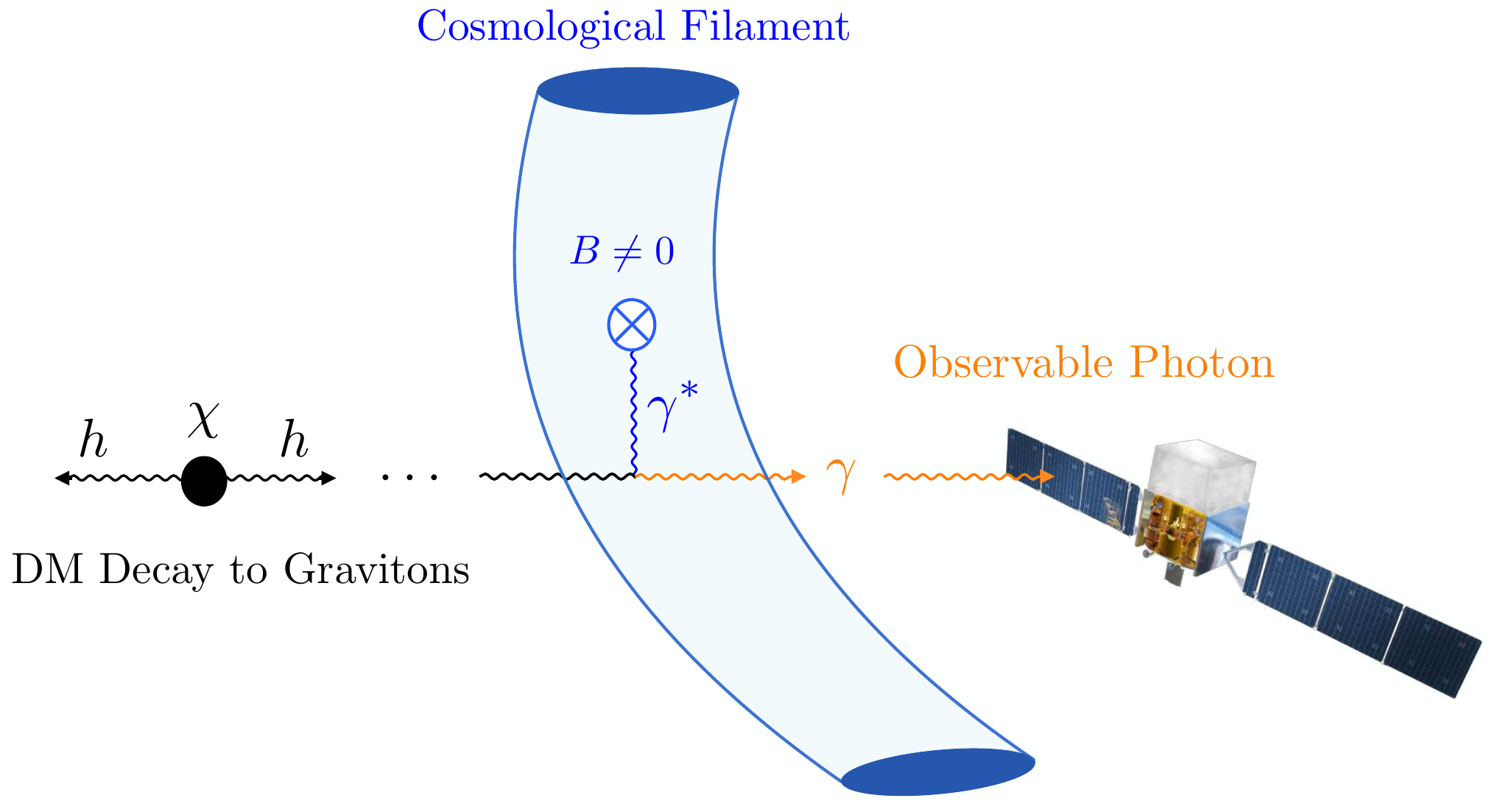}
\vspace{-0.cm}~
\caption{
Schematic cartoon of our search strategy: a DM particle $\chi$ decays to gravitons at early times. The gravitons propagate through magnetized cosmological filaments and can oscillate into photons along the line-of-sight towards an observer. This flux is an irreducible prediction of any DM model that allows for decays to gravitons. 
 }
\label{fig:cartoon}
\end{figure}

However, gravitons are not completely dark. As they traverse spatial regions with nonzero magnetic fields, they can convert to photons through a SM process known as the Gertsenshtein effect \cite{Gertsenshtein,Boccaletti:1970pxw}.
Furthermore, our universe provides ample opportunities for such conversion events in a variety of astrophysical and cosmological settings.  
Since the length scale for graviton-photon conversion typically exceeds the Hubble radius, the most promising sources for realizing this effect are cosmological filaments, which feature $\sim$ 100 nG magnetic fields on long $\sim$ Mpc coherence lengths \cite{Amaral:2021mly}, and fill an order unity fraction of our universe's volume \cite{Vernstrom:2021hru}.

\begin{figure*}
\includegraphics[width=  
\linewidth]{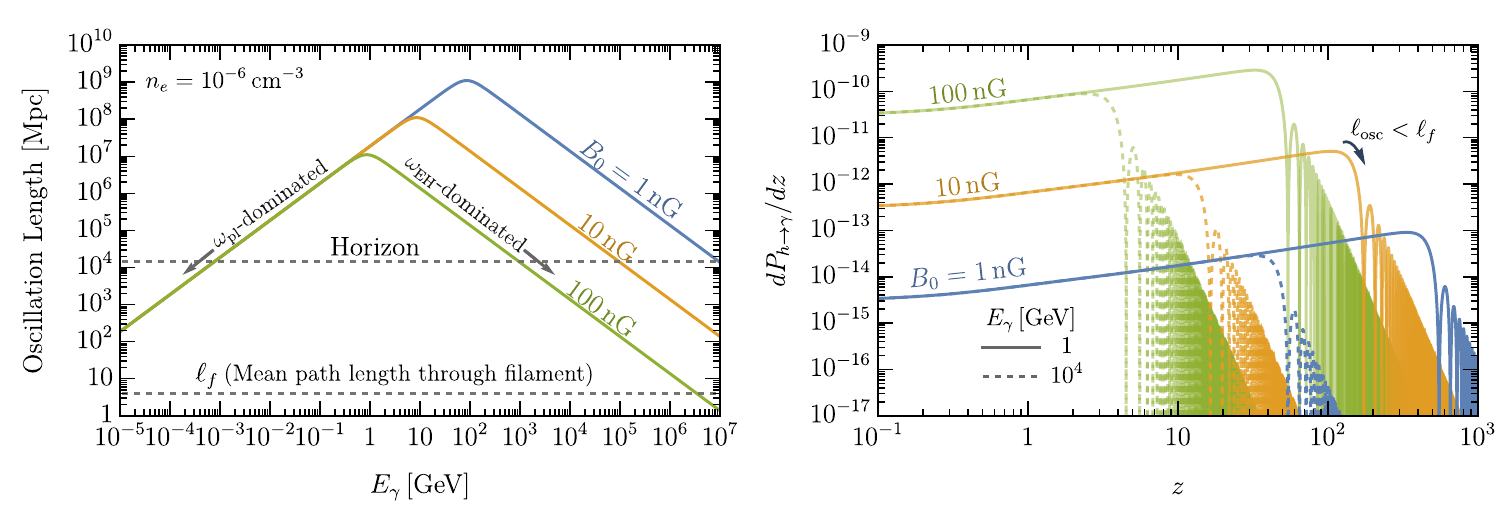
} 
\vspace{-0.5cm}
\caption{
{\bf Left:} The present-day oscillation length for graviton-photon conversion from \Eq{eq:losc} for various choices of $B_0$. Note that for most graviton energies of interest in this paper (10 GeV$-$TeV), $\omega_{\rm EH} \gg \omega_{\rm pl}$ which is to the right of the peak. The upper horizontal black curve is the proper distance of the present-day cosmological horizon $\sim 3.3 H_0^{-1}  \approx 14.4$ Gpc. The lower horizontal curve is the typical path length traveled by a graviton through a single filament today, $\approx 4$ Mpc.
}
\label{fig:lOscDiagram}
\end{figure*}

In this \emph{Letter}, we introduce a new search strategy for dark matter decays to gravitons, which leverages the Gertsenshtein effect in magnetized cosmological filaments. If DM particles decay to gravitons in the early universe, the latter traverse many filament patches and have a small, but appreciable probability of converting to {\it photons} along the line of sight to an observer. For a given choice of dark matter mass and lifetime, these Gertsenshtein photons add an irreducible contribution to the isotropic gamma-ray background (IGRB), and we set the first-ever limits on the graviton decay channel based on this technique. Over a wide range of DM masses, we set world-leading limits on the DM lifetime if this decay channel predominates.

\medskip
\noindent
{\bf Graviton Decays.} 
The lowest dimension operators that allow a spin-0 field $\chi$ to decay to gravitons $h$ have non-minimal\footnote{Naively there are several simpler operators that might induce $\chi \to hh$ decays -- for example, $\sqrt{g} \chi R $ 
$\sqrt{g} \chi R^2,$ or $ \sqrt{g} \chi R^{\mu\nu} R_{\mu\nu}$. However, these bilinear interactions with gravitons either vanish in flat space or in the true vacuum of the theory \cite{Alonzo-Artiles:2021mym,Ema:2021fdz}. } \cite{Alonzo-Artiles:2021mym,Ema:2021fdz, Landini:2025jgj}  quadratic couplings to the Riemann tensor $R_{\mu\nu\alpha\beta}$ and its dual $\tilde R_{\mu\nu\alpha\beta}$: 
\be
\label{eq:gravity}
{\mathscr{L} }_{\chi hh} =  \sqrt{|g|} \left( \frac{\chi}{\Lambda}  R^{\mu\nu\alpha\beta} R_{\mu\nu\alpha\beta} + 
 \frac{\chi}{\tilde \Lambda}R^{\mu\nu\alpha\beta} \tilde R_{\mu\nu\alpha\beta}  \right),~~
\ee
 where $\Lambda$ and $\tilde\Lambda$ are scales associated with integrating out heavy particles and 
  $|g|$ is the metric determinant. The $\chi$ lifetime arising from either operator is \cite{Ema:2021fdz}
 \be
    \label{eq:fidicualLifetime}
     \tau_\chi
     = \frac{4\pi \Lambda^2 M_{\rm Pl}^4}{m_\chi^7} 
     \approx 2 \times 10^{20} \, {\rm s} \left(\frac{10^6 \, \rm GeV}{m_\chi} \right)^{5}  \!\left(\frac{\Lambda}{m_\chi} \right)^{2} \!\! ,~~
 \ee 
 where $M_{\rm Pl}$ is the reduced Planck mass.
Alternatively, $\chi$ may decay to a single graviton and another light dark state, in which case the  lifetime can  scale as only $\propto M_{\rm Pl}^2$ and may be much shorter than the estimate in Eq.\,\eqref{eq:fidicualLifetime}.

 In our treatment, we thus remain agnostic about the origin of the operators contributing to DM decays to gravitons. For our purposes, it suffices to specify the  lifetime for $\chi$, which is equivalent  to freely choosing a set of coefficients for these operators; throughout this analysis, we assume that a suitable choice can always be made to realize any $\chi$ lifetime, subject only to observational bounds.

Although we are agnostic about the mechanism that populates $\chi$ in the early universe, there are many viable possibilities that are consistent with our scenario. Over a wide range of $\chi$ masses, the abundance can arise through gravitational interactions during inflation \cite{Markkanen:2015xuw,Ema:2015dka,Ford:1986sy}, reheating \cite{Chung:1998zb}, radiation domination \cite{Garny:2015sjg,Landini:2025jgj}, or primordial black hole evaporation \cite{Matsas:1998zm,Fujita:2014hha}. More generally, the abundance can arise through any production mechanism compatible with the existence of the operators in Eq. \eqref{eq:gravity}, as long as $\chi \to hh$ remains the dominant decay channel. 

\medskip
\noindent
{\bf Gertsenshtein Effect.}
In this section we review the Gertsenshtein effect whereby a graviton converts to 
a photon in the presence of a uniform magnetic field. Following the formalism in Ref. \cite{Domcke:2020yzq}, the $h \to \gamma$ conversion
probability over some distance $\ell$ can
be written as
\begin{align}
\label{eq:gerg}
    P_{h \to \gamma} = |K_{\rm osc}|^2 \ell_{\rm osc}^2 \sin^2 \left ( \frac{\ell}{\ell_{\rm osc} } \right)~,
\end{align}
with the characteristic oscillation length
\be
\label{eq:losc}
    \ell_{\rm osc} =  2\left[\omega^2(1-\mu)^2 + \kappa^2 B^2\right]^{-1/2},
\ee
where  $\omega$ is the incident graviton/photon energy, $\kappa^2 = 16 \pi G$, and
 $B$ is the angle-averaged transverse magnetic field strength. We have also defined an oscillation amplitude $K_{\rm osc}$ and refractive index $\mu$, which respectively satisfy 
\be
    |K_{\rm osc}| = \frac{\sqrt{\mu} \kappa B}{1+ \mu} ~~,~~
\mu = \sqrt{1- (\omega_{\rm pl}^2 - \omega_{\rm EH}^2)/\omega^2}~~,
\ee
where characteristic plasma frequency is 
\be
\label{eq:wPL}
\omega_{\rm pl} = \sqrt{ \frac{ 4\pi \alpha  n_e}{m_e}} \sim 10^{-14} \, {\rm eV} \brac{n_e}{10^{-6} \rm cm^{-3}}^{1/2} ~,
\ee
and the Euler-Heisenberg frequency is 
\be
\label{eq:wEH}
\omega_{\rm EH} \approx  \sqrt{ \frac{7 \alpha }{45\pi} } \brac{B}{B_{c} } \omega 
\sim  10^{-14} \, {\rm eV} \brac{B}{ 60 \rm \, nG} \brac{\omega}{\rm GeV}\!,~~~~~
\ee
where $\alpha$ is the fine structure constant, $B_{c} = m_e^2/e \approx 4.4 \times 10^{13}$ G \cite{Raffelt:1987im}, and $n_e$ the ambient electron number density.

\begin{figure*}[t!]
\includegraphics[width =0.98\columnwidth]{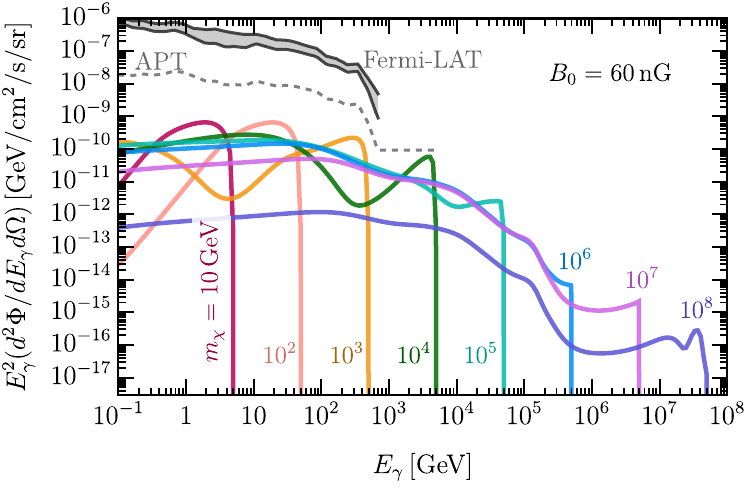}~~~     ~
\includegraphics[width =0.98\columnwidth]{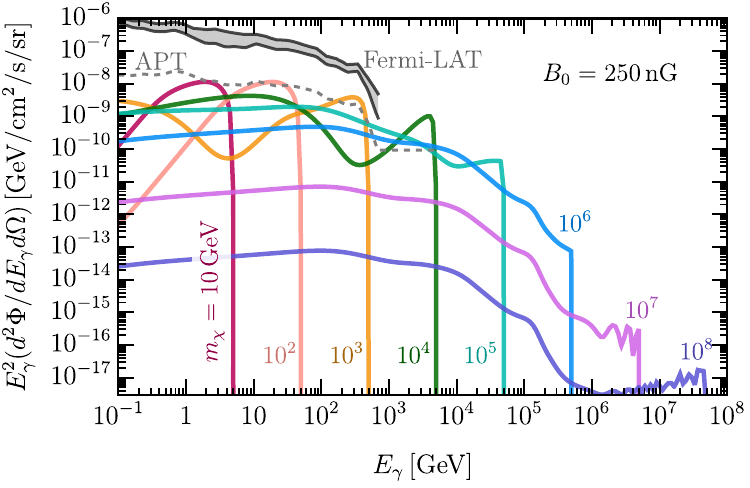}
\caption{
 {\bf Left:} Differential photon fluxes from $\chi \to hh$ decays followed by graviton-photon conversion 
 in cosmological filaments. Each curve  represents
 a different choice of DM mass in GeV assuming the same values of $B_0 = 60$ nG, $\tau_\chi = 10^{19}$s, $\ell_f = 4$ Mpc, and $f_{\rm vol} = 0.15$. 
 {\bf Right}: Same as left, but with $B = $ 250 nG. 
 }
\label{fig:flux}
\end{figure*}

\medskip
\noindent
{\bf Cosmological Filaments.}
Cosmological filaments provide an irreducible source of magnetic fields
that can be modeled 
  as \cite{Carretti:2024bcf}
 \be
B(z) = B_0 (1+z)^2,
 \ee
 where $z$ is the cosmological redshift and $B_0$ is the average present day field value, which ranges between $1-600$ nG \cite{Vernstrom:2021hru, Brown:2017dwx, Vacca:2018rta, OSullivan:2018shr, Vernstrom:2019gjr, OSullivan:2020pll, Locatelli:2021byc, Amaral:2021mly, Carretti:2022tbj, Hoang:2023uyx, Anderson:2024giz, Carretti:2024bcf} with coherence lengths that range between 1-10 Mpc \cite{Vernstrom:2019gjr, Amaral:2021mly,Locatelli:2021byc}.  It is estimated that $10-30$\% of our present Hubble volume consists of such filaments \cite{Libeskind:2017tun,Dome:2023ska}. 

Assuming characteristic present-day filament parameters 
 $B_0 = 60$ nG 
 and $n_e = 10^{-6} \, \rm{cm}^{-3}$ 
in Eqs. \eqref{eq:wPL} and \eqref{eq:wEH}, the plasma and Euler-Heisenberg frequencies are
 are comparable for $\omega \sim {\rm GeV} $ in the gamma-ray band. For these reference values, the $\omega$ dependence in \Eq{eq:losc} is negligible, so 
 the present-day oscillation length for $h\to\gamma$ conversion is simply
 \be
 \ell_{\rm osc} \approx 2\times 10^4 \, {\rm Gpc}, 
 \ee
 which vastly exceeds the present day Hubble radius of $H_0^{-1} \approx 4 \, {\rm Gpc}$. Indeed, Fig. \ref{fig:lOscDiagram} shows that $\ell_{\rm osc}$ is much larger than the horizon or typical filament coherence length for a wide range of photon energies.\footnote{In contrast, the oscillation length of gravitational waves from astrophysical or primordial sources is typically in the extreme subhorizon since the graviton energies are much smaller than those we consider in this work.}
Thus, we treat the conversion probability in each filament as an independent trial
and a typical graviton will encounter many such patches while propagating over $\sim$ Gpc distance scales.

Within a single filament, a graviton has a photon conversion probability of 
\begin{align}
\label{eq:probSingleEncounter}
P^{(1)}_{h\to \gamma}(z)   &= |K_{\rm osc}(z)|^2 \ell_{\rm osc}(z)^2 \sin^2 \left[ \frac{\,\ell_f(z)}{\ell_{\rm osc}(z) } \right] 
\\
&\approx 4\pi G B(z)^2  \ell_f(z)^2 ~,
\end{align}
where  $\ell_f(z)$ is the length traversed by a graviton within a single filament at redshift $z$. In the second line, we write the approximate form of $P_{h\rightarrow \gamma}^{(1)}$ when $\ell_f \ll l_{\rm osc}$ and $\mu \simeq 1$, which is often the case.  Dirac showed \cite{dirac1943approximate,sanchez2004use} that the isotropically averaged chord length of a convex shape of volume $V$ and area $A$ is $4V/A$,
implying that for filaments, roughly cylinders of radius $r$ and height $h > r$,
\begin{align}
    \label{eq:ellfDef}
    \langle \ell_f(z) \rangle  \approx \frac{2r(z)}{1+ r(z)/h(z)} \approx \frac{2 r_0}{1+z} ,
\end{align}
and the average squared chord length \cite{case1953introduction}
\begin{align}
    \langle \ell_f(z)^2\rangle \approx \frac{16r(z)^2}{3} \simeq [2.3 r(z)]^2  \, .
\end{align}
Simulations indicate that the typical filament radius today is $r_0 \approx 1-3$ Mpc \cite{colberg2005intercluster, Wang:2024qej}. 
In \Eq{eq:ellfDef}, we assume the filament radius scales with redshift as $(1+z)^{-1}$, though the exact redshift scaling is not crucial since most of the converted photon flux is generated at low redshifts. For the remainder of this work, we adopt the fiducial value $r_0 = 2$ Mpc. 

In a small physical distance $dr$, the graviton traverses a number of filaments 
\begin{align}
\label{eq:dNf}
    dN_{f} = \frac{ f_{\rm vol}(z) }{\,\, \ell_f(z)} dr  = 
    \frac{ f_{\rm vol}(z) }{\, \,\ell_f(z)}
    \frac{dz}{ H(z)(1+z)} \, ,
\end{align}
where $f_{\rm vol}(z) \approx 0.15$ is  the volume-filling fraction of cosmological filaments at redshift $z$ \cite{Libeskind:2017tun,Dome:2023ska}. Thus, the cumulative probability for graviton-photon conversion between two arbitrary redshifts is
\begin{align}
    P_{h\to \gamma} = \int_{z_1}^{z_2} dN_{f}(z) P^{(1)}_{h\to \gamma}(z)  \, \, ,
\end{align}
and we can write the  differential  conversion probability per unit redshift
\be
\label{eq:conversion-prob}
    \frac{dP_{h\to \gamma}}{dz} =
    \frac{dN_{f} }{dz}
     \frac{dP_{h\to \gamma}}{dN_{f}}  = 
\frac{ f_{\rm vol}(z) }{\, \,\ell_f(z)}
    \frac{P_{h\to \gamma}(z)}{ H(z)(1+z)}~,~~
\ee
where we have used \Eq{eq:dNf}. In principle, $f_{\rm vol}(z)$ may depend on redshift, but here we assume this quantity to be a constant, given that the converted photon flux mainly arises at low redshift.
In our treatment, we also conservatively assume that graviton propagation outside of filaments does not contribute to the conversion probability.

\begin{figure*}[t!]
\hspace{-0.2cm}
\includegraphics[width =1.5\columnwidth]{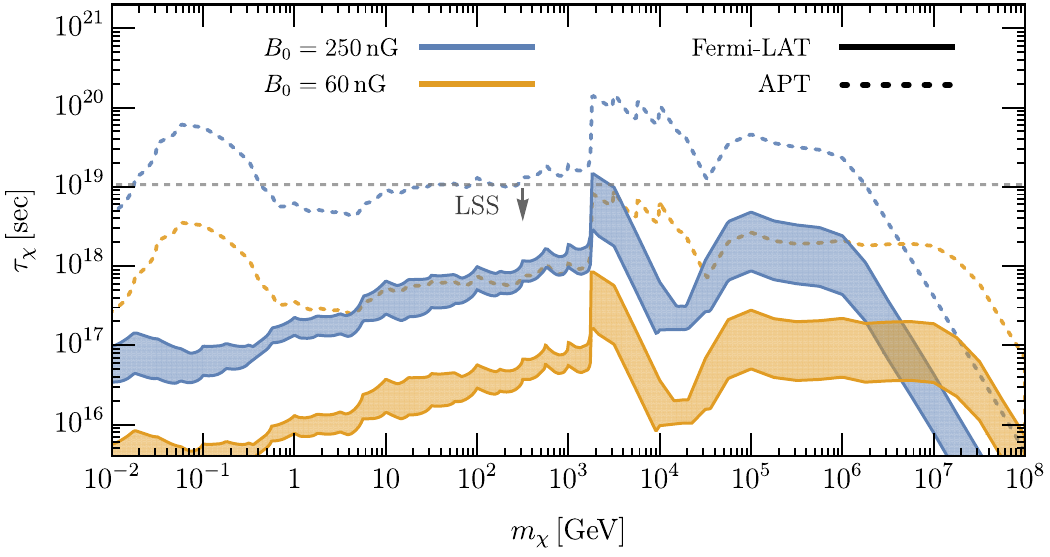}
\caption{
Limits on the DM lifetime assuming $\chi \to hh$ decays, where $h$ is a graviton,  followed by $h\to \gamma$ conversion in cosmic filaments. Each panel assumes filaments with coherence length $\ell_f =4$ Mpc and volume fraction $f_{\rm vol} = 0.15$. The horizontal dotted line is the constraint on DM decaying to dark radiation from CMB and large scale structure observations, which allow for $\sim 4\%$ of DM to have decayed invisibly on cosmological timescales at 95 \% confidence \cite{Poulin:2016nat}. Note, however, that this limit assumes the $\Lambda$CDM cosmological model and the precise bound on the lifetime can vary in alternative cosmological scenarios. }
 \label{fig:constraintPlot}
\end{figure*}
Nevertheless, gravitons from DM decays can also convert to photons inside the magnetic fields found in galaxy clusters, and voids. Since the graviton energies we consider are large ($\omega \gg 1$ GeV), the graviton-photon oscillation length is typically far larger than the coherence sizes of these systems (see Fig. \,\ref{fig:lOscDiagram}). As a result, Eqns.\,\eqref{eq:probSingleEncounter} and \eqref{eq:dNf} indicate that to optimize the total probability of $h \rightarrow \gamma$ conversion, the product of $B^2 \ell f_{\rm vol}$ should be maximized, where $B$, $\ell$ and $f_{\rm vol}$ are the typical magnetic fields, scale sizes, and universe filling fraction of these systems, as shown in Table\,\ref{tab:BFieldScalings}.
 
In this analysis, we conservatively focus on filaments as the most promising regions for studying the contribution of our DM model to diffuse emission. Including contributions from clusters and voids would further strengthen our results. Targeted observations\footnote{Here, ``target searches'' refer to observations where the telescope points at a single target, while ``diffuse emission searches'' refer to cases in which the cumulative emission from objects contributes to the IGRB or other types of diffuse emission.} will also be promising for this kind of study. In particular, for targeted observations, the total probability of conversion is of the form of Eq.\,\eqref{eq:conversion-prob}, but with $f_{\rm vol} \rightarrow 1$ and the integration only over the physical extent of the target. An additional advantage is that the point-source sensitivity of telescopes is typically orders of magnitude better than their sensitivity to diffuse emission, particularly for gamma-ray telescopes. This analysis focuses on contributions to diffuse emission, while a systematic study of targeted observations is deferred to future work.
\begin{table}[]
    \centering
\begin{tabular}{|l|c|c|c|c|}
    \hline 
    \multicolumn{1}{|c|}{Domain} & 
    \multicolumn{1}{c|}{$B$ (nG)} & 
    \multicolumn{1}{c|}{$\ell$ (Mpc)} & 
    \multicolumn{1}{c|}{$f_{\rm vol}$} & 
    \multicolumn{1}{c|}{~$B^2 \ell f_{\rm vol}$~} 
    \\
    \hline
    Filaments  &  $10\!-\!100$  &  $\sim 1$ & $\sim 10^{-1}$ & $10\!-\!10^{3}$ \\
    Clusters &  $\sim 10^3$ & $\sim 10^{-2}$ & $\sim 10^{-2}$ & $\sim 10^{2}$ \\
    Voids &  $ 10^{-7}\!-\!10^{-4}$ & $\sim 1$ & $\sim 1$ & $10^{-14}\!-\!10^{-8}$ \\
    \hline
\end{tabular}

    \caption{Comparison of cosmological domains where graviton to photon conversion can occur. Filaments \cite{Amaral:2021mly,Vernstrom:2021hru,Libeskind:2017tun}, galaxy clusters \cite{Libeskind:2017tun,Pfrommer:2009hn}, and voids \cite{Libeskind:2017tun,Taylor:2011bn,Rodriguez-Medrano:2023uen} possess different typical magnetic fields $B$, coherence lengths $\ell$, and volume filling fractions $f_{\rm vol}$. The contribution to the diffuse gamma-ray background from DM decays to gravitons scales with $B^2 \ell f_{\rm vol}$, which is largest for cosmic filaments.}
    \label{tab:BFieldScalings}
\end{table}
\medskip

\noindent
{\bf Photon Flux.}
The flux of gravitons from DM decays before the redshift of graviton-photon conversion $z_c$ can be written
\begin{align}
     \label{eq:gravitonFlux}
    \frac{d \Phi_h}{dE}(z_c) =
    \frac{
    \Omega_{\chi} \rho_c}{\tau_\chi m_{\chi} }
    \int_{z_c}^\infty  \!\! \frac{dz}{H(z)} \frac{dN_h}{dE}(z) \, ,
\end{align}
where $\tau_\chi$ is the DM lifetime, $\Omega_\chi = 0.24$ is the DM energy fraction,  $\rho_c$ is the critical density, and 
\begin{align}
\label{eq:dNdE}
    \frac{dN_h}{dE}(z) = 2 \delta \! \left[ E(1+z) - \frac{ m_\chi}{2} \right] \, ,
\end{align}
 is the  present-day graviton energy spectrum due to decays at redshift $z$, where $E(1+z)$ is the graviton energy at the time of decay, and $E$ is the present-day graviton (or photon) energy. In the late universe, the Hubble rate in \Eq{eq:energyFlux} can be written
\be
H(z) = H_0 \sqrt{ \Omega_m(1+z)^3 + \Omega_\Lambda}~,
\ee
where $H_0 = 68$ km$\,$s$^{-1}$Mpc$^{-1}$ is the Hubble constant, $\Omega_\Lambda  = 0.69$ is the present-day dark energy  fraction, $\Omega_m = 0.31$ is the present-day matter fraction  \cite{Planck:2018vyg}.

To derive the total photon flux at redshift $z$ from all prior conversion events in cosmological filaments, we convolve \Eq{eq:gravitonFlux} with
\Eq{eq:conversion-prob} to obtain
\begin{align}
    \label{eq:energyFlux}
    \frac{d \Phi_\gamma}{dE}(z) = \int_{z}^\infty dz_c \frac{dP_{h \rightarrow \gamma}(z_c)}{dz_c}   \frac{d\Phi_h}{dE}(z_c)  \,  .
\end{align}
Using the delta function from \Eq{eq:dNdE} to eliminate the integral in \Eq{eq:gravitonFlux}, the 
photon flux from Eq.\,\eqref{eq:energyFlux} can be written
\begin{align}
\label{eq:photon-flux}
    \frac{d \Phi_\gamma}{dE }(z) =  
    \frac{2 \,\Omega_{\chi} \rho_c }{\tau_\chi m_{\chi}  E  H(z_*) }
    \int_{z}^{z_*} \! dz_c
    \frac{dP_{h \rightarrow \gamma}(z_c)}{dz_c}  
  ~,
\end{align}
where we have defined 
$
z_* \equiv (m_\chi - 2 E)/(2E)
$
as the redshift of $\chi$ decay and $z_* > z$.

The spectrum in \Eq{eq:photon-flux} represents the photon flux at redshift $z$ before accounting for the electromagnetic cascades that result from photon interactions with their surroundings. 
As they traverse cosmological distances, these photons scatter cosmic microwave background and extragalactic background photons to undergo pair production
via $\gamma \gamma \to e^+e^-$ reactions, which attenuate photons above $\sim$ 100 GeV energies. The $e^+e^-$ pairs produced in these reactions inject softer secondary photons via $e^+e^- \to \gamma \gamma$ annihilation, $e \gamma \to e\gamma$ inverse-Compton scattering, and synchrotron radiation.
Collectively, these cascades redistribute photon energies and asymptote to a nearly universal secondary spectrum for input energies above the TeV scale \cite{Blanco:2018esa}. To model this secondary photon population, we use the injected flux from \Eq{eq:photon-flux} as an input for the $\gamma$-Cascade V4 code package \cite{Capanema:2024nwe}, which generates the present-day photon flux.

\medskip
\noindent
{\bf Results.}
Fig.\,\ref{fig:flux} shows representative photon fluxes from DM decays to gravitons followed by graviton-photon conversion in filaments, with each colored contour corresponding to a different DM mass.  For $E_\gamma \gtrsim 1$ TeV, cascades are important and DM masses between $10^4 \lesssim m_\chi \lesssim 10^6$ GeV possess a near universal spectrum at sub-100 GeV energies. For $m_\chi \gtrsim 10^6$ GeV, the Euler-Heisenberg contribution to the photon mass is sufficiently large to drop $\ell_{\rm osc}$ below $\ell_f$ (see Fig. \ref{fig:lOscDiagram}), and the probability of conversion is suppressed due to the $\sin^2(\ell_f/\ell_{\rm osc})$ term in Eq.\,\eqref{eq:probSingleEncounter} becoming important. This oscillation in the $P_{h \rightarrow \gamma}$ leads to an imprinted oscillatory pattern at the highest photon energies in the spectrum and the reduction in amplitude relative to the universal spectrum for low-energy photons.

The gray band of Fig.\,\ref{fig:flux} represents the observed IGRB flux from Fermi-LAT telescope data \cite{DiMauro:2016cbj}. In the future, the Advanced Particle-astrophysics Telescope (APT) will be able to probe the $\gamma$-ray sky with an order of magnitude improvement in $\gamma$-ray sensitivity and effective area compared to Fermi-LAT \cite{Alnussirat:2021tlo}. The APT will be able to resolve many of the currently unresolved point sources that contribute to the IGRB observed by Fermi-LAT.

In terms of the acceptance $\xi$, which is provided in Fig. 3 of Ref. \cite{Alnussirat:2021tlo},
 the APT sensitivity to the IGRB flux is expected to increase between a factor of $\sqrt{\xi_{\rm APT}/\xi_{\rm Fermi}}$
 and $\xi_{\rm APT}/\xi_{\rm Fermi}$
 if the measurement is background or signal dominated, respectively; these ratios are approximately $\sqrt{10}$ ($10$) for photon energies above $100$ GeV and $\sqrt{500}$ ($500$) for photon energies around $10$ MeV. 
Although the acceptance in Ref. \cite{Alnussirat:2021tlo} is only plotted out to TeV energies, the APT team anticipates that the telescope will be able to measure photons up to energies of 5 TeV, so we extrapolate the APT acceptance out to this higher value\footnote{Private communication with Jim Buckley.} as shown by the dashed gray $\text{APT}$ line of Fig. \ref{fig:flux}.

To constrain the DM lifetime, we compare candidate photon spectra from $\chi \to hh$ decays and place limits on parameter values for which this graviton induced photon flux exceeds IGRB observations.
In Fig. \ref{fig:constraintPlot}, we present our main result: the limit on the DM lifetime assuming $\chi \to hh$ decays with 100\% branching fraction over a wide range of masses. The thickness of this band reflects the uncertainty of the Fermi-LAT IGRB measurement, as shown in the gray band of Fig. \ref{fig:flux}. For DM masses near the TeV scale, our analysis sets the strongest limits on the DM lifetime assuming a reference value of $B_0 = 250$ nG. The sharp rise in the bounds around $m_\chi \approx 2$ TeV is due to the steep drop in the IGRB near $E_\gamma \approx 1$ TeV coinciding with the sharp cutoff of the photon spectrum at $E_\gamma = m_\chi/2$.  Fig.\, \ref{fig:constraintPlot}  also  shows  generic bounds on DM decays to invisible final states from Ref. \cite{Poulin:2016nat} (gray dashed curve) and APT sensitivity projections assuming an acceptance enhancement factor of $\xi_{\rm APT}/\xi_{\rm Fermi}$ to reduce the IGRB flux from point sources. 

\medskip
\noindent
{\bf Discussion.}
In this \emph{Letter}, we have placed the first indirect-detection limits on dark matter decaying to gravitons by exploiting the Gertsenshtein effect to calculate the flux of photons from graviton conversion in cosmological filaments. For a given DM mass and lifetime, this photon population adds an irreducible component to  the extragalactic gamma-ray background. Using this predicted photon flux, we conservatively set world leading limits on the DM lifetime across a wide range of masses, assuming a 100\% branching fraction to graviton pairs. 
The results of our paper are based on the state-of-the-art knowledge in the literature regarding the average filament length, number density, volume-filling fraction, and magnetic field strengths within these cosmic bridges. We emphasize that the Gertsenshtein effect is a Standard Model processs and that the only new physics considered here is the hypothesis that dark matter can decay to gravitons. 

Approximately half of the universe's baryons are believed to reside in cosmic filaments in the form of a warm-hot diffuse gas, known as  the warm-hot intergalactic medium (WHIM). The thermal emission and synchrotron radiation from the WHIM in the filaments result in X-rays and radio waves. 
Therefore, future observations in the X-ray band with AXIS \cite{Reynolds:2023vvf}, Athena \cite{Barret:2019qaw} and XRISM \cite{XRISMScienceTeam:2020rvx}, as well as through 
radio observations with ngVLA \cite{ngVLA}, SKA \cite{Weltman:2018zrl} and its pathfinders, will make it possible to further constraint the properties of filaments.  

While waiting for new observations, hydrodynamical simulations such as EAGLE~\cite{Schaye:2014tpa, Crain:2015poa, EAGLE:2017}, BAHAMAS \cite{McCarthy_2016}, Illustris-TNG \cite{Pillepich:2017jle, Springel:2017tpz}, MillenniumTNG \cite{Pakmor:2022yyn}, and FLAMINGO~\cite{Schaye:2023jqv, Kugel:2023wte} are particularly well-suited for studying these structures due to their large box sizes (100 Mpc for EAGLE and BAHAMAS, 205 Mpc for TNG300, 500 Mpc for MillenniumTNG, 1–2.8 Gpc for FLAMINGO), which enable the analysis of extensive inter-cluster filaments—ideal environments for this type of dark matter search.

Intriguingly, given the cosmologically long distances required for appreciable graviton-photon conversion in filaments, this signal arises predominantly from extragalactic sources, not from the Galactic center, as one typically finds for DM decays to all other SM particle species. Thus, a smoking gun for this scenario is a statistically significant excess in the extragalactic gamma-ray flux, with negligible flux from the galactic center. 
 
{\bf Acknowledgments.}
We thank Carlos Blanco, Nassim Bozorgnia, Jim Buckley, Dan Carney, Cyril Creque-Sarbinowski, Neal Dalal, Chris Dessert, Josh Foster, 
Shy Genel, Nagisa Hiroshima, Dan ``Danimal" Hooper, Junwu Huang, Mehr U Nisa, Lara Arielle Phillips, Nick Rodd, Ken Van Tilburg, and Evan Vienneau for helpful conversations. 
EP is grateful for the hospitality of Perimeter Institute where part of this work was carried out. Research at Perimeter Institute is supported in part by the Government of Canada through the Department of Innovation, Science and Economic Development and by the Province of Ontario through the Ministry of Colleges and Universities. DD is supported by the James Arthur Postdoctoral Fellowship and is grateful to the Fermi National Accelerator Laboratory where part of this work was carried out. This manuscript has been authored in part by  Fermi Forward Discovery Group, LLC under Contract No. 89243024CSC000002 with the U.S. Department of Energy, Office of Science, Office of High Energy Physics. 

\bibliographystyle{utphys3}
\bibliography{biblio}

\end{document}